\documentclass{article}
\usepackage{amssymb}
\usepackage{graphicx}
\usepackage{epstopdf}
\usepackage{amsfonts}
\usepackage{amsmath}
\newtheorem{theorem}{Theorem}
\newtheorem{acknowledgement}[theorem]{Acknowledgement}

\begin{document}

\title{Garrett approximation for asymmetric rectangular potentials and its
applications to quantum well infrared photodetectors}
\author{Victor Barsan \\
IFIN-HH, 30 Reactorului\\
and \\
UNESCO\ Chair of HHF, 407 Atomistilor\\
077125 Magurele, Romania}
\maketitle

\begin{abstract}
Garrett's approximation for the calculation of bound states energy in square
wells is applied in a consistent way. So, one obtains simple formulas for
these energies, with errors of about $1\%$ for moderate, and $0.01\%$ for
deep wells. The application in of our results in the physics of quantum well
infrared photodetectors is briefly discussed.
\end{abstract}

\section{Introduction}

The symmetric square well (Fig. 1), and more complicated rectangular wells
(Figs. 2, 3, 4), are popular models of semiconductor heterojunctions
potentials. The wave functions of particles moving in such potentials can be
easily expressed in terms of elementary functions, but the bound states
energy eigenvalues are given by transcendental equations, which defies exact
solutions. However, several approximate analytic formulas of the eigenenergy
were proposed. One of them, sometimes called the Barker approximation \cite%
{[Barker]}, has been sucessfully used to explain, \textit{inter alia}, the
optical absorbtion spectra of certain quantum wells / heterojunctions \cite%
{[Biwas]}. Powerful mathematical softs, giving accurate numerical results
for the energy levels or any other quantity of interest, are available, but
the attempt of finding an analytical solution, even an approximate one,
remains appealing.

An interesting approach for obtaining analytical approximate expressions for
the bound state energy of a particle in a square well was proposed by
Garrett \cite{[Garrett]}. Unlike other approximations, obtained by various
mathematical tricks applied to the transcendental eigenvalue equations \cite%
{[Barker]}, \cite{[dABG]}, \cite{[VB-RD]},\ \cite{[VB-RRP]}, Garrett's
method is based on a simple physical idea: as the main difference between
the infinite and finite square well is the fact that, in the first case, the
wall is impenetrable, but in the second one, the wave function penetrates
the wall on a certain distance $\delta $, the energy of a bound state in a
finite well of length $L$ should be satisfactorily approximated by the
energy of the corresponding bound state, in an infinite well of length $%
L+\delta $. Garrett also mentioned that the same approach could be applied
to asymmetric wells. However, he did not applied sistematically this idea,
not even for a symmetric square well. This simple exercise has been done
recently in \cite{[VB-RD]}, \cite{[VB-EJP]}, when it has been also shown
that the Garrett approximation, obtained by a consequent application of
Garrett's idea, is equivalent to Barker's approximation for deep wells.

As already mentioned, the square wells are not only a useful framework for
discussing undergraduate problems of one-dimensional quantum mechanics: they
have also important practical applications. One of them refers to the
physics of semiconductor heterojunctions, characterized by various types of
square wells. For the design and production of a class of quantum wells
infrared photodetectors, it is essential to describe correctly the bound
state energy in such piecewised defined potentials. We shall explain how
Garrett's approximation, in conjunction with other approximate analytical
methods, can give quite accurately the energy levels of interest in the
study of these devices.

The structure of this papers is the following: in Section 2, we obtain the
Garrett approximation of the energy of bound states, in symmetric and
asymmetric wells. In Section 3 we give the equation of the dimensionless
wave vector of the bound states of an asymmetric well, mainly in order to
estimate the errors of the Garrett approximation, for both symmetric and
asymmetric wells. In Section 4, we show how the Garrett approximation can
give simple and relatively accurate expressions of the energy levels
relevant for the physics of quantum well infrared photodetectors. The last
section is devoted to conclusions.

\section{Garrett's approximation for finite square wells}

In order to expose Garrett's approach, let us mention that the energy levels
of a particle of mass $m$\ in an infinite rectangular well of length $L$\ is
given by the well-known formula:

\begin{equation}
E_{n}^{\left( 0\right) }=n^{2}\frac{\pi ^{2}\hbar ^{2}}{2mL^{2}}  \tag{1}
\label{1}
\end{equation}

The same particle, moving in a finite square well of depth $V$ and the same
length $L$ (see Fig.1),\ can penetrate the wall of the well on a distance of
about:

\begin{equation}
\delta ^{\left( 1\right) }=\frac{\hbar }{\left[ 2m\left( V-E_{n}^{\left(
0\right) }\right) \right] ^{1/2}}  \tag{2}  \label{2}
\end{equation}%
So, it behaves similarly to a particle moving in an infinite well of length $%
L+2\delta ^{\left( 1\right) };$ consequently, its energy can be written as:

\begin{equation}
E_{n}^{\left( 1\right) }=n^{2}\frac{\pi ^{2}\hbar ^{2}}{2m\left( L+2\delta
^{\left( 1\right) }\right) ^{2}}  \tag{3}  \label{3}
\end{equation}%
which can be considered a first correction to (\ref{1}). Using $%
E_{n}^{\left( 1\right) },$ we can define a new characteristic length:

\begin{equation}
\delta ^{\left( 2\right) }=\frac{\hbar }{\left[ 2m\left( V-E_{n}^{\left(
1\right) }\right) \right] ^{1/2}}=\frac{\hbar \left( L+2\delta ^{\left(
1\right) }\right) }{\left[ 2mV\left( L+2\delta ^{\left( 1\right) }\right)
^{2}-\pi ^{2}\hbar ^{2}n^{2}\right] }  \tag{4}  \label{4}
\end{equation}%
and we can define, similar to (\ref{2}), a second correction to (\ref{1}). A
consistent application of this approach requests an infinity of steps, the
general one being:

\begin{equation}
\delta ^{\left( q+1\right) }=\frac{\hbar \left( L+2\delta ^{\left( q\right)
}\right) }{\left[ 2mV\left( L+2\delta ^{\left( q\right) }\right) ^{2}-\pi
^{2}\hbar ^{2}n^{2}\right] }  \tag{5}  \label{5}
\end{equation}%
Putting:

\begin{equation}
\lim_{q\rightarrow \infty }\delta ^{\left( q\right) }=\Delta ,\ y=\frac{%
2\Delta }{L}\   \tag{6}  \label{6}
\end{equation}%
and taking the limit $q\rightarrow \infty $\ in the both sides of (6), we
obtain a quartic equation in $y:$

\begin{equation}
4P^{2}y^{4}+8P^{2}y^{3}+\left( 4P^{2}-\pi ^{2}n^{2}-4\right) y^{2}-8y-4=0 
\tag{7}  \label{7}
\end{equation}%
where $P$ is the strength of the finite well:

\begin{equation}
P=\sqrt{2mV}\frac{L}{2\hbar }  \tag{8}  \label{8}
\end{equation}%
So, $P$ is a dimensionless quantity, characterizing both the potential $%
\left( L,\ V\right) $\ and the particle $\left( m\right) $.

If the well is large, the cubic and quartic terms in eq. (\ref{7}) can be
neglected, and one obtains:

\begin{equation}
y\simeq \frac{1}{P}+\frac{1}{P^{2}},\ \ \Delta =\frac{L}{2}\left( \frac{1}{P}%
+\frac{1}{P^{2}}\right)   \tag{9}  \label{9}
\end{equation}%
and the energy of the $n$-th bound state in the finite well can be
approximated by:

\begin{equation}
E_{n}=n^{2}\frac{\pi ^{2}\hbar ^{2}}{2m\left( L+2\Delta \right) ^{2}}=n^{2}%
\frac{\pi ^{2}\hbar ^{2}}{2mL^{2}\left( 1+\frac{1}{P}+\frac{1}{P^{2}}\right)
^{2}}  \tag{10}  \label{10}
\end{equation}

Also, as $2mE_{n}=\hbar ^{2}k_{n}^{2}$:

\begin{equation}
Lk_{n}=\frac{\pi n}{1+\frac{1}{P}+\frac{1}{P^{2}}}  \tag{11}  \label{11}
\end{equation}%
So, in this approximation, the quantity $Lk_{n},$ called sometimes
dimensionless wave vector, depends linearly on $n.$

Let us remind that the equation giving the exact values of the wavectors $%
k_{n}$ is \cite{[LL]}, \cite{[Messiah]}:

\begin{equation}
n\pi -kL=2\arcsin \frac{kL}{P}  \tag{12}  \label{12}
\end{equation}%
A plot of the bound state energies given by the solutions of the exact
equation (\ref{12}) and by the Garrett approximation (\ref{10}) is given in
Fig. 5.

For a simple asymmetric square well (see Fig. 2), we expect, for the same
physical reasons as for a symmetric one, that, instead of (\ref{11}), the
dimensionless wave vector will be given by:

\begin{equation}
Lk_{n}=\frac{\pi n}{1+\frac{1}{2}\left( \frac{1}{P_{1}}+\frac{1}{P_{1}^{2}}%
\right) +\frac{1}{2}\left( \frac{1}{P_{2}}+\frac{1}{P_{2}^{2}}\right) } 
\tag{13}  \label{13}
\end{equation}%
where we put:

\begin{equation}
P_{i}=\sqrt{2mV_{i}}\frac{L}{2\hbar },\ \ i=1,2\   \tag{14}  \label{14}
\end{equation}

We shall call formulas (\ref{11}), (\ref{12}) Garrett approximations for the
symmetric, respectively asymmetric wells. Even if Garrett did not obtain
these formulas explicitely, he gave the intuitive idea of this new and
simple approximation. It is more convenient to use the formulas for the wave
vector than that for energy, for instance (\ref{11}) instead of (\ref{10}),
as they are linear in $n.$ For deep wells, $P\gtrsim 10,$ the Garrett
approximation is equivalent to Barker's one \cite{[Barker]}.

The ground state dimensionless wave vector $Lk_{1}$ of the symmetric square
well, given by the "exact" solution of the eigenvalue equation (disks) and
by the Garrett approximation (squares), for strength $P=1,2,...,15$, are
plotted in Fig. 6. For shallow wells $\left( P\leq 3\right) ,$ the
approximation is poor (about $10\%)\,$, but for moderate $\left( 4\leq P\leq
9\right) $ and deep $\left( 10\leq P\right) $\ wells it is of the order $%
10^{-2}$, respectively $10^{-3}$.

\section{The simple asymmetric well}

Let as consider the simplest generalization of the symetric well (Fig. 2),
called sometimes symple asymmetric square well. Its corresponding
Schroedinger equation:

\begin{equation}
\left( -\frac{\hbar ^{2}}{2m}\frac{d^{2}}{dx^{2}}+V\left( x\right) \right)
\psi =E\psi   \tag{15}  \label{15}
\end{equation}%
can be written simpler, as

\begin{equation}
\psi ^{\prime \prime }+\left[ k^{2}-U\left( x\right) \right] \psi =0\  
\tag{16}  \label{16}
\end{equation}%
if we introduce $U\left( x\right) $ instead of $V\left( x\right) $\ by:

\begin{equation}
V\left( x\right) =\frac{\hbar ^{2}}{2m}U\left( x\right) .  \tag{17}
\label{17}
\end{equation}%
We shall define, following Messiah \cite{[Messiah]}, Ch. III, \S 6 (see also 
\cite{[LL]}, \S 22, problem 2)

\begin{equation}
U\left( x\right) =U_{3}\Theta\left( b-x\right) +U_{2}\Theta\left( x-b\right) \Theta\left( a-x\right) +U_{1}\Theta\left( x-2\right)   \tag{18}  \label{18}
\end{equation}

The bound state wave functions have the form:

\begin{equation}
\psi \left( x\right) =\left\{ 
\begin{array}{c}
A_{1}e^{-K_{1}x},\ x>a \\ 
A_{2}\sin \left( kx+\varphi \right) ,\ b<a<a \\ 
A_{3}e^{K_{3}x},\ x<b%
\end{array}%
\right.   \tag{19}  \label{19}
\end{equation}

\bigskip We shall put:

\begin{equation}
k=\sqrt{k^{2}-U_{2}},\ K_{1}=\sqrt{U_{1}-k^{2}},\ K_{3}=\sqrt{U_{3}-k^{2}} 
\tag{20}  \label{20}
\end{equation}

Without restricting the generality, we can choose $U_{2}=0$ and define:

\begin{equation}
L=b-a,\ U=U_{1},\ U_{3}=\left( 1+\delta \right) U,\ \delta >0  \tag{21}
\label{21}
\end{equation}

Coming back to the general case, we get:

\begin{equation}
K_{1}=\sqrt{U-k^{2}},\ K_{3}=\sqrt{\left( 1+\delta \right) U-k^{2}}  \tag{22}
\label{22}
\end{equation}

The eigenvalue equation associated to the solution (\ref{19}) has the form:

\begin{equation}
n\pi -Lk=\arcsin \frac{Lk}{2P\sqrt{\left( 1+\delta \right) }}+\arcsin \frac{%
Lk}{2P}  \tag{23}  \label{23}
\end{equation}%
For a symmetric well, $\delta =0$ and it becomes:

\begin{equation}
n\pi -Lk=2\arcsin \frac{Lk}{2P}  \tag{24}  \label{24}
\end{equation}

The relative errors of the Garrett approximation of the dimensionless wave
vectors $Lk$ whose exact value are given by the eqs. (\ref{23}), (\ref{24})
are plotted in Fig. 7; the approximation is more precise for an asymmetric
well than for a symmetric one with the same strength, and decreases if the
asymmetry increases. The explanation of this behaviour is simple.
Physically, an asymmetric wall is less penetrable than a symmetric one, of
the same strength; in other wods, it produces a stronger confinement effect.
Mathematically, the function entering in the r.h.s. of (\ref{24}), $\arcsin
x,$ exists for $x<1;$ the function entering in the r.h.s. of (\ref{23}), $%
\arcsin x+\arcsin \frac{x}{\sqrt{1+\delta }},$ exists for a shorter
interval, $x<\left( 1+\delta \right) ^{-1/2}.$ As 
\begin{equation*}
\arcsin x=x+\frac{1}{6}x^{3}+O\left( x^{5}\right) ,
\end{equation*}%
a shorter interval favorizes the linear approximations - in our case,
Garrett approximation. This is illustrated by Fig. 8, which gives the
graphical solutions of the eqs. (\ref{23}), (\ref{24}). It is clear that,
for deep wells, the approximate solution is very close to the exact one. The
accuracy of the approximation is also favorized by the fact that its error
changes its sign for bound states situated in the upper half of the well.

\begin{figure}
\begin{center}
\includegraphics[width=\textwidth]{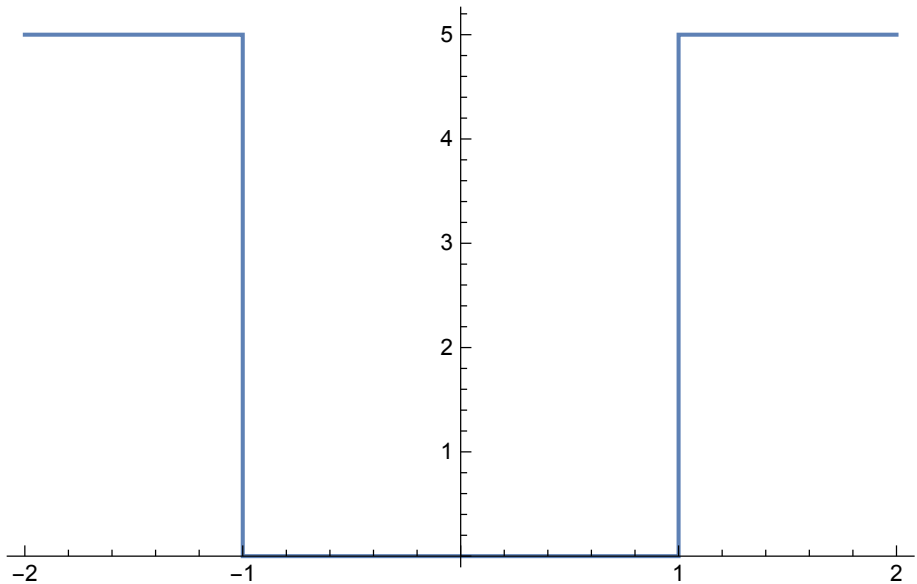}
\end{center}
\caption{ Symmetric square well}
\end{figure}

\begin{figure}
\begin{center}
\includegraphics[width=\textwidth]{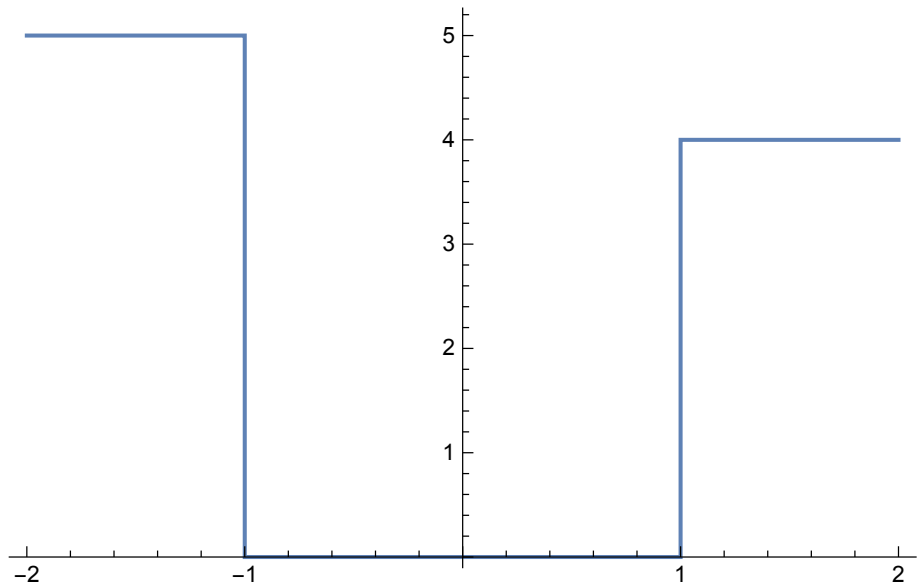}
\end{center}
\caption{ Simple aymmetric square well }
\end{figure}

\begin{figure}
\begin{center}
\includegraphics[width=\textwidth]{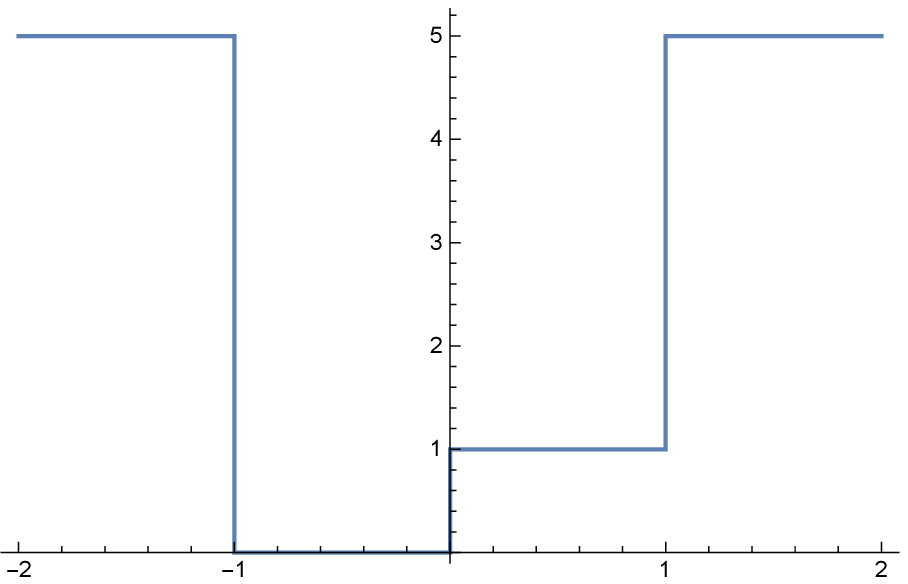}
\end{center}
\caption{ Stepped square well with symmetric barriers }
\end{figure}

\begin{figure}
\begin{center}
\includegraphics[width=\textwidth]{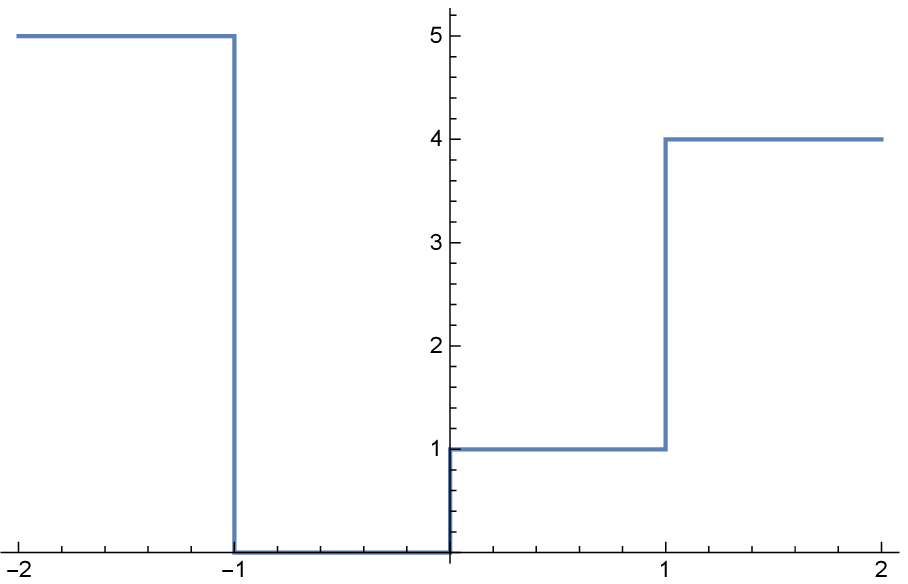}
\end{center}
\caption{ Stepped square well with asymmetric barriers }
\end{figure}

\begin{figure}
\begin{center}
\includegraphics[width=\textwidth]{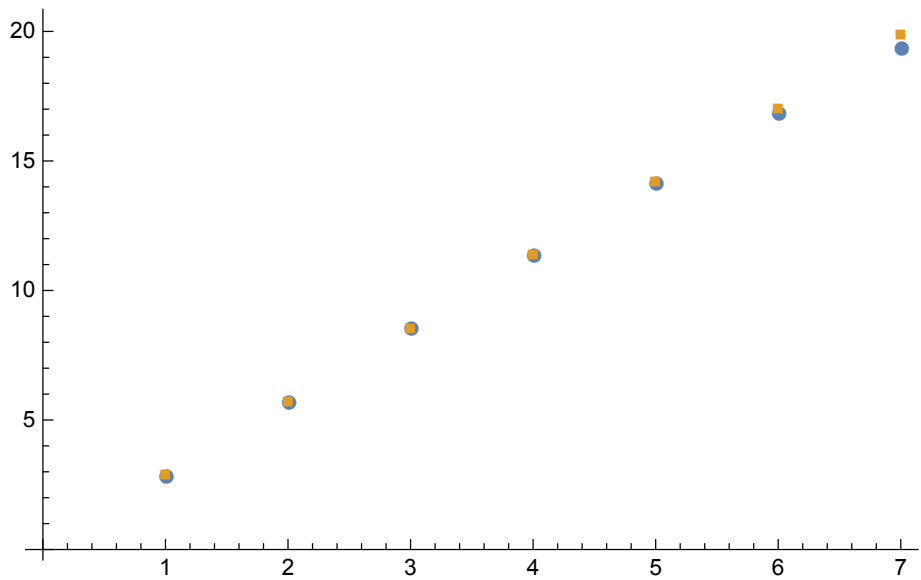}
\end{center}
\caption{Dimensionless bound states wave vectors for a symmetric square well with 𝑃=10, given by the "exact" solution of the eigenvalue equation (disks) and by the Garrett approximation (squares}
\end{figure}

\begin{figure}
\begin{center}
\includegraphics[width=\textwidth]{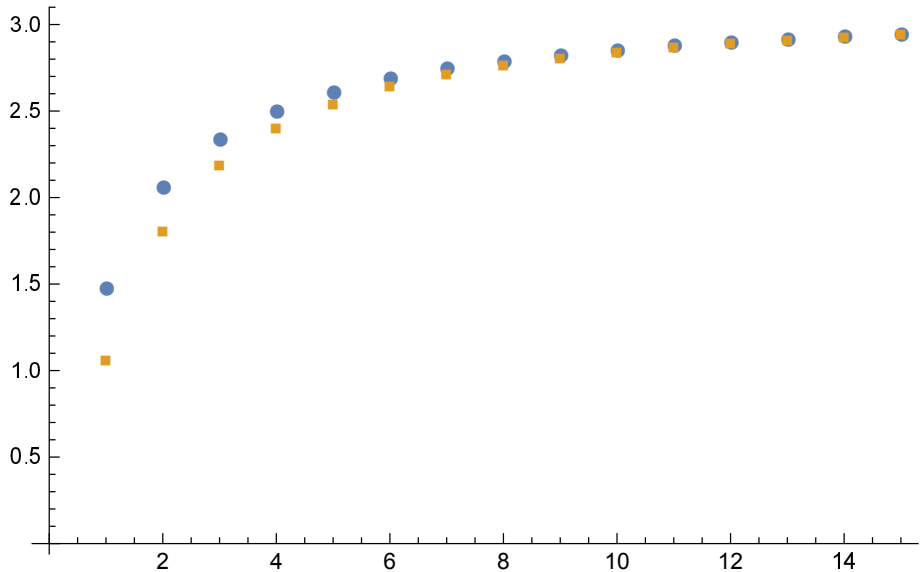}
\end{center}
\caption{ Dimensionless ground state wave vector of the symmetric square well, given by the "exact" solution of the eigenvalue equation (disks) and by the Garrett approximation (squares), for P=1, 2, ...15.}
\end{figure}

\begin{figure}
\begin{center}
\includegraphics[width=\textwidth]{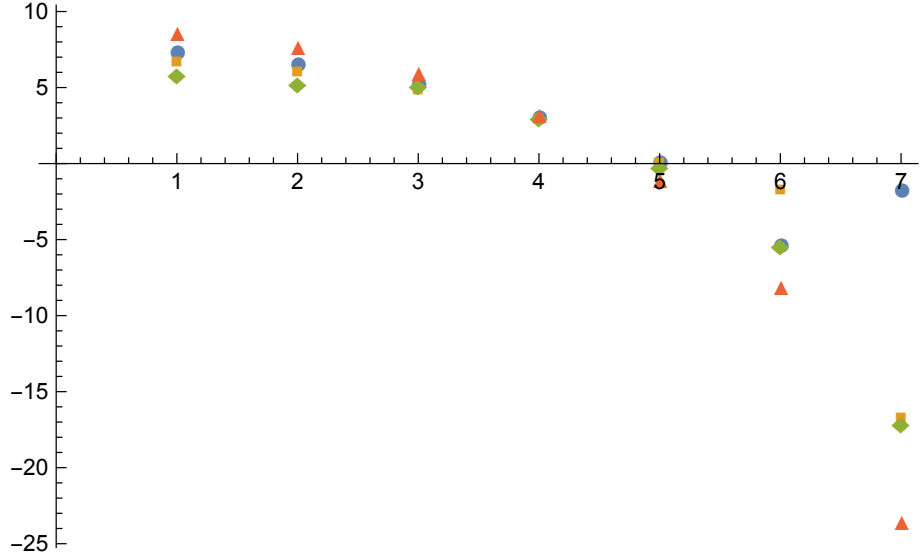}
\end{center}
\caption{ Error of Garrett′𝑠 approximation, in units of 0.1 percent, for the dimensionless bound state wave vectors of a simple asymmetric well with 𝑃=10 and delta=1/2 (disks), delta=1 (squares, edges paralell to axes) and delta=2 (squares, diagonals paralell to axes); also, for delta=0 (triangles).}
\end{figure}

\begin{figure}
\begin{center}
\includegraphics[width=\textwidth]{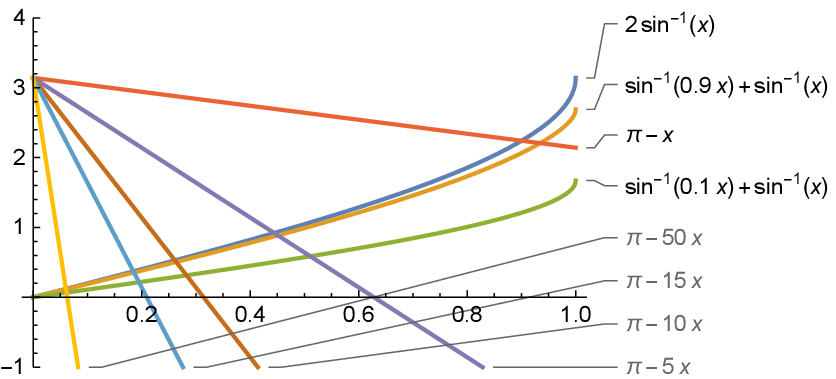}
\end{center}
\caption{ Graphical solutions of the eq.(23).}
\end{figure}

\section{Applications to quantum well infrared photodetectors (QWIP)}

The use of symmetric QW as photodetectors is hampered by the fact that, due
to the symmetry of wave functions - in other words, to the existence of even
and odd states - the selection rules forbid some photoabsorbtion processes.
If the QW is asymmetric, such selection rules are no longer active, and the
photoabsorption is enhanced \cite{[Gupta]}. In the last two decades or so,
much work has been done on the effects of the asymmetry on optical and
electrical properties of the detectors. Of course, the greatest interest is
connected to infrared detectors, with important applications in imaging,
pollution monitoring, spectroscopy of genes, etc. \cite{[Dupont]}.

Such applications were studied by Brandel et al. \cite{[Brandel]}, whose
infrared detector is composed by thin quantum wells $\left( L=20A\right) \ $%
of pure and dopped $GaAs$, sandwiched between thick $Al_{x}Ga_{1-x}As$
barriers $\left( L_{b}=300A\right) $. More exactly, the QW is composed by
two thin films, one of $GaAs$ and the other of $Al_{y}Ga_{1-y}As,$ $y<x.$ As
the films have slightly different compositions, there is a certain differnce
between the energy offsets, and they form a stepped asymmetric well (\cite%
{[Brandel]}, Fig. 1), with enhanced photodetctivity. For a well with $L=40A$
and $V=1eV,\ $the potential strength is $P\simeq 5,$ (we approximated the
effective electron mass inside the well with the free electron mass)
characterizing a moderately deep well, when the error of Garrett
approximation is $\sim 1\%$. As $1eV$ is the energy corresponding to an
infrared photon, $\lambda =1,24\mu ,$ and interband transitions involve
energies of a fraction of $eV,$ photodetectors similar to those described in 
\cite{[Brandel]} can be used in the near and mid-infrared region of spectra,
and their characteristics can be calculated with reasonable accuracy usung
Garrett's approximation.

Also, in this case, the penetration length is of the order

\begin{equation*}
\Delta =\frac{a}{2P}=4A, 
\end{equation*}%
much smaller than the dimension of the barrier $\left( L_{b}=300A\right) ;$\
the overlap of wave functions of neighbor wells is negligible, so the wells
can be considered as decoupled.

If the relative dopping of the thin films inside the barrier is small, $y\ll
1,$ which happends when the main role of the step is to create asymmetry, so
to inhance the photon absorbtion, the first-order perturbation theory to the
levels of a symmetric well, given by very precise analytical approximations,
will give errors of the order $y$ for the energy levels of the stepped well.
Indeed, in a slightly dopped $GaAs$ asymmetric stepped quantum well, the
band discontinuity is $\Delta =0.8y$ (in eV) \cite{[Dupont]}. A similar
asymmetric well characterizes the device studied by Dupont et al. \cite%
{[Dupont]}, for generating coherent terahertz waves.

Hostut et al. \cite{[Hostut]} studied a tri-colour infrared photodetector,
obtained by repeated layers of three $GaAs$\ QW units. In each unit, the
wells are thin $GaAs$ layers $\left( \sim 50A\right) $ separated by thick $%
Al_{x}Ga_{1-x}As$ barriers $\left( L_{b}=300A\right) .$ The wells are simple
asymmetric wells, with a small asymmetry, see Fig. 1 of \cite{[Hostut]}. For
this type of photodetector, the transition from the ground state of the well
to the highest state is of interest; if the parameters of the detector are
adapted to near- or mid-infrared region, the wells are quite deep $\left(
P\sim 10\right) ,$ so Garrett approximation would be appropriate for
calculation of the ground state energy.

\section{Conclusions}

Following in a consistent manner Garrett's idea to approximate the bound
states energy of a finite square well by the energy of the corresponding
state of a conveniently defined infinite well, we obtain simple formula for
the levels of symmetric and asymmetric wells, with errors of about $1\%$ for
moderate, and about $0.1\%$ for deep wells. In this way, we can calculate,
with reasonable approximation, the energy levels of QW, relevant for the
physics of quantum well infrared photodetectors. It is quite surprising that
such a simple approach works in a problem governed by complicated and
untractable eigenvalue equations. The results can be applied to stepped
wells, using elementary perturbation theory of quantum mechanics, at least
in the case of small concentrations of dopants.

\begin{acknowledgement}
The author acknowledges the financial support of the ANCSI - IFIN-HH project
PN 18 09 01 01/2018.
\end{acknowledgement}

\end{document}